\documentclass[sigconf]{acmart}
\usepackage{acmart-taps}

\usepackage{amsmath,amssymb,amsfonts}
\usepackage{comment}
\usepackage{eso-pic}

\newcommand{\SSDBMstatusheader}{%
  \AddToShipoutPictureFG*{%
    \AtPageUpperLeft{%
      \raisebox{-0.55in}[0pt][0pt]{%
        \makebox[\paperwidth][c]{%
          \small\itshape
          Accepted and presented as a poster at the 38th International Conference on Scalable Scientific Data Management (SSDBM 2026), San Diego, California, August 12--13, 2026.%
        }%
      }%
    }%
  }%
}
\AtBeginDocument{%
  }

\setcopyright{acmlicensed}
\copyrightyear{2026}
\acmYear{2026}
\acmDOI{XXXXXXX.XXXXXXX}
\acmConference[Conference acronym SSDBM]{38th International Conference on Scalable Scientific Data Management}{August 12--13,
  2026}{San Diego, California}

\renewcommand\footnotetextcopyrightpermission[1]{}

\begin{document}

\title{Utility-Driven Spatial Data Sampling for UAV-Assisted Scientific Smart Farming}

\author{Keiwan Soltani}

\orcid{0000-0002-6938-196X}
\authornotemark[1]
\affiliation{%
  \institution{Missouri University of Science and Technology}
  \city{Rolla}
  \state{Missouri}
  \country{USA}
}
\email{ksoltani@mst.edu}

\author{Sajal K. Das}
\affiliation{%
  \institution{Missouri University of Science and Technology}
  \city{Rolla}
  \state{Missouri}
  \country{USA}
  }
\email{sdas@mst.edu}

\renewcommand{\shortauthors}{Soltani et al.}

\begin{abstract}
Large smart-farming deployments generate continuous scientific data from spatially distributed sensors, including soil, humidity, temperature, crop-health, and pest-related measurements. In vast agricultural fields, however, an energy-constrained unmanned aerial vehicle (UAV) often cannot collect data from every sensor during each mission. Existing UAV-assisted collection methods typically optimize coverage, route length, data volume, or freshness, but they do not always distinguish between data that is merely available and data that is scientifically valuable. This poster introduces a utility-driven spatial sampling framework for UAV-assisted smart farming. The field is partitioned into grid cells, each sized according to the UAV’s ground coverage range. After an initial exploration phase, each cell receives a scientific utility score based on freshness, redundancy, anomaly likelihood, and model uncertainty. The UAV then selects and visits a subset of high-utility cells under battery and return-to-base constraints. The proposed framework reframes UAV-based collection as adaptive scientific data management rather than exhaustive sensing.
\end{abstract}

\SSDBMstatusheader
\maketitle

\section{Introduction}
Smart farms generate continuous soil, environmental, crop-health, and pest-related data from spatially distributed sensors~\cite{soltani-smartfarm}. In large and connectivity-limited farms, an energy-constrained UAV cannot collect all sensor data during every mission~\cite{soltani-data2024}. Existing UAV-assisted collection methods commonly optimize trajectory length, coverage, or freshness, but these objectives do not distinguish between data that are merely available and data that are scientifically informative~\cite{UAV-agriculture2024}. We propose a utility-driven spatial sampling framework that assigns each UAV-coverage cell a value based on freshness, anomaly likelihood, model uncertainty, and redundancy. The UAV then selects an energy-feasible subset of high-value cells rather than exhaustively collecting data from the entire field.

\section{System Overview}
The agricultural field $\mathcal{A}$ contains a set of $N$ static sensor nodes $S= \{s_1, s_2, \dots, s_N\}$ . Due to the vast field area, the UAV is unable to collect data from all sensors in one round of flight. The field is therefore divided into grid cells, where the size of each cell corresponds to the UAV’s ground coverage area. When the UAV flies to the center of a cell, it can collect data from all sensors located inside that cell using Orthogonal Frequency-Division Multiple Access (OFDMA) mechanism, Simultaneously. 

The framework has two phases. In the initial exploration phase, the UAV collects data from all grid cells in one or more early flights. This phase builds a baseline spatial profile of the field, including normal sensor behavior, temporal patterns, spatial correlation among cells, historical anomaly levels, and model performance across regions. In the adaptive collection phase, the UAV no longer visits all cells. Instead, each cell is assigned a scientific utility score, and the UAV plans an energy-feasible route over a selected subset of cells.

This design is especially suitable for scientific data management because the system maintains a cell-level metadata table. For each cell, the table stores the last collection time, recent summaries, correlation with nearby cells, anomaly history, model uncertainty, estimated data volume, and expected collection energy. The UAV’s mission is then planned using this metadata before departure and can be updated after each mission.
\vspace{-0.1in}
\section{Cell-Level Scientific Utility}
Let $C=\{c_1,c_2,\cdots,c_M\}$ denote the set of grid cells where $M$ is the number of total cells $M\leq N$. Each cell $c_i$ contains one or more sensors and is associated with a utility score $\mathcal{S}_i$ that estimates how valuable it is to collect data from that cell in the current mission. This scores combines four core dimensions, Freshness, Redundancy, anomaly and model uncertainty.

\textbf{Freshness} captures how outdated the cell's data are. If $t$ is the current time and $t^{l}_i$ is the last time cell $c_i$ was collected, then a simple freshness need can be defined as
\vspace{-0.05in}
\begin{equation}
    F_i= \frac{1}{|c_i|}\sum_{j\in c_i} \left[1- e^{-\lambda(t-t^{l}_j)} \right],
\end{equation}
where $t^{l}_{j}$ is the last time UAV collected data from sensor $j$ in cell $c_i$ and $\lambda$ controls how quickly the data become stale. A larger $F_i$ indicates that the cell should be revisited.

For each sensing modality $m\in \mathcal{M}$, such as soil moisture, temperature, humidity, or water level, the measurements of sensors located in cell $c_i$ are first aggregated into a representative cell-level time series as
\vspace{-0.1in}
\begin{equation}
    \hat{x}_i^{m}(t) = \frac{1}{|c_i|}\sum_{s_{j}\in c_i}X_{i,j}^m(t),
\end{equation}
where $X_{i,j}^m(t)$ denotes the reading of sensor $s_j$ in cell $c_i$ from modality $m$ at time $t$.

\textbf{Redundancy} captures whether the information provided by a cell is already represented by nearby cells. Let $\mathcal{N}_i$ denote the neighboring cells of $c_i$. For each sensing modality, the redundancy of Cell $c_i$ is estimated by the average positive historical correlation between its aggregated time series and those of its neighboring cells:
\vspace{-0.1in}
\begin{equation}
    R_i^m = \frac{1}{|\mathcal{N}_i|}\sum_{c_j \in \mathcal{N}_i} \max \left(0.corr(\hat{x}_i^m, \hat{x}_j^m) \right),
\end{equation}
\begin{equation}\nonumber
   R_i=\sum_{m\in \mathcal{M}} \omega_m R_i^m, \quad\quad \sum_{m\in \mathcal{M}} \omega_m =1
\end{equation}

A high $R_i$ indicates that Cell $c_i$ exhibits temporal behavior similar to nearby cells; therefore, its information may be inferred from neighboring observations and may be less urgent to collect.

\textbf{Anomaly likelihood }captures whether the recent condition of a cell differs substantially from its own historical behavior. For each modality, let $\mu_i^{m}$ and $\sigma_i^{m}$ denote the historical mean and standard deviation of the aggregated cell-level series $\hat{x}^{m}_i$. The modality-specific anomaly score is defined as:
\vspace{-0.1in}
\begin{equation}
    A_i^m = \left| \frac{\hat{x}_i^m(t) - \mu_i^m}{\sigma_i^m + \epsilon} \right|
\end{equation}
where $\epsilon$ is a small constant that avoids division by zero. The anomaly likelihood of Cell $c_i$ is then obtained by combining the modality-specific anomaly scores:
\vspace{-0.05in}
\begin{equation}
    A_i =\sum_{m\in \mathcal{M}}\Phi_m A_i^m, \quad \quad \sum_{m\in \mathcal{M}} \Phi_m=1
\end{equation}

A high $A_i$ indicates that the current aggregated measurements of the cell are unusual relative to their historical pattern, which may reflect localized water stress, irrigation irregularity, environmental change, or another agricultural event.

\textbf{Model uncertainty} captures whether collecting new data from this cell may improve a downstream learning model, such as a pest-detection or crop-health classifier. A predictive model receives the recent aggregated measurements and temporal features of Cell $c_i$ and produces a probability distribution over $K$ agricultural-condition classes $p_i=[p_{i,1},p_{i,2}, \dots, p_{i,k}]$. The uncertainty of the prediction is measured using normalized Shannon entropy:
\vspace{-0.08in}
\begin{equation}
    U_i = -\frac{1}{\log K}\sum_{k=1}^K p_{i,k} \log p_{i,k}.
\end{equation}

A high $U_i$ indicates that the model cannot confidently determine the condition of Cell $c_i$. Therefore, collecting new data from this cell may be particularly useful for updating the downstream agricultural monitoring model.
Finally, the overall scientific utility score for cell $c_i$ is then defined as 
\vspace{-0.1in}
\begin{equation}
    \mathcal{S}_i= \alpha F_i + \beta A_i + \gamma U_i - \delta R_i,
\end{equation}
where $\alpha, \beta, \gamma, \delta$ are application-specific weights. This formulation reward cells that are stale, anomalous, or uncertain, while penalizing the cells whose data are likely redundant.

\section{Energy-Aware Route Selection}
Once cell scores are computed, the UAV must decide which cells to visit and in what order. This is not a standard Traveling Salesman Problem because the UAV does not need to visit all cells. Instead, the UAV must select a subset of cells that maximizes total collected utility while satisfying its battery limit. The problem is closer to an orienteering problem, where each candidate location has a reward and the route is constrained by a travel budget.

Let $\mathcal{R} \subseteq C$ be the ordered set of cells selected for a UAV mission. The objective is
\vspace{-0.1in}
\begin{equation}
    \max_{\mathcal{R} \subseteq C}\sum_{c_i\in R} \mathcal{S}_i,
\end{equation}
subject to
\vspace{-0.1in}
\begin{equation}
    E_t(\mathcal{R}) + E_h(\mathcal{R}) + E_c(\mathcal{R}) + E_r(\mathcal{R}) \leq B,
\end{equation}

where $B$ is the UAV battery budget. The term $E_t(\mathcal{R})$ represents the energy required to travel along the selected route, $E_{h}(\mathcal{R}) $is the energy spent hovering at selected cell centers, $E_{c}(\mathcal{R})$ is the energy required for communication and data reception, and $E_{r}(\mathcal{R})$ guarantees that the UAV has enough remaining energy to return safely to the base station. 

A practical heuristic can select the next cell using marginal utility per additional energy:
\vspace{-0.1in}
\begin{equation}
    MU(c_i) = \frac{\mathcal{S}_i}{\Delta E_{route}(c_i)},
\end{equation}
where $\Delta E_{route}(c_i)$ s the extra energy required to insert cell $c_i$ into the current route. A cell is feasible only if the UAV can visit it, collect data, and still return to the base station:
\vspace{-0.1in}
\section{Preliminary Evaluation}
\begin{figure}
\centerline{\includegraphics[width=\columnwidth,height=6cm,keepaspectratio]{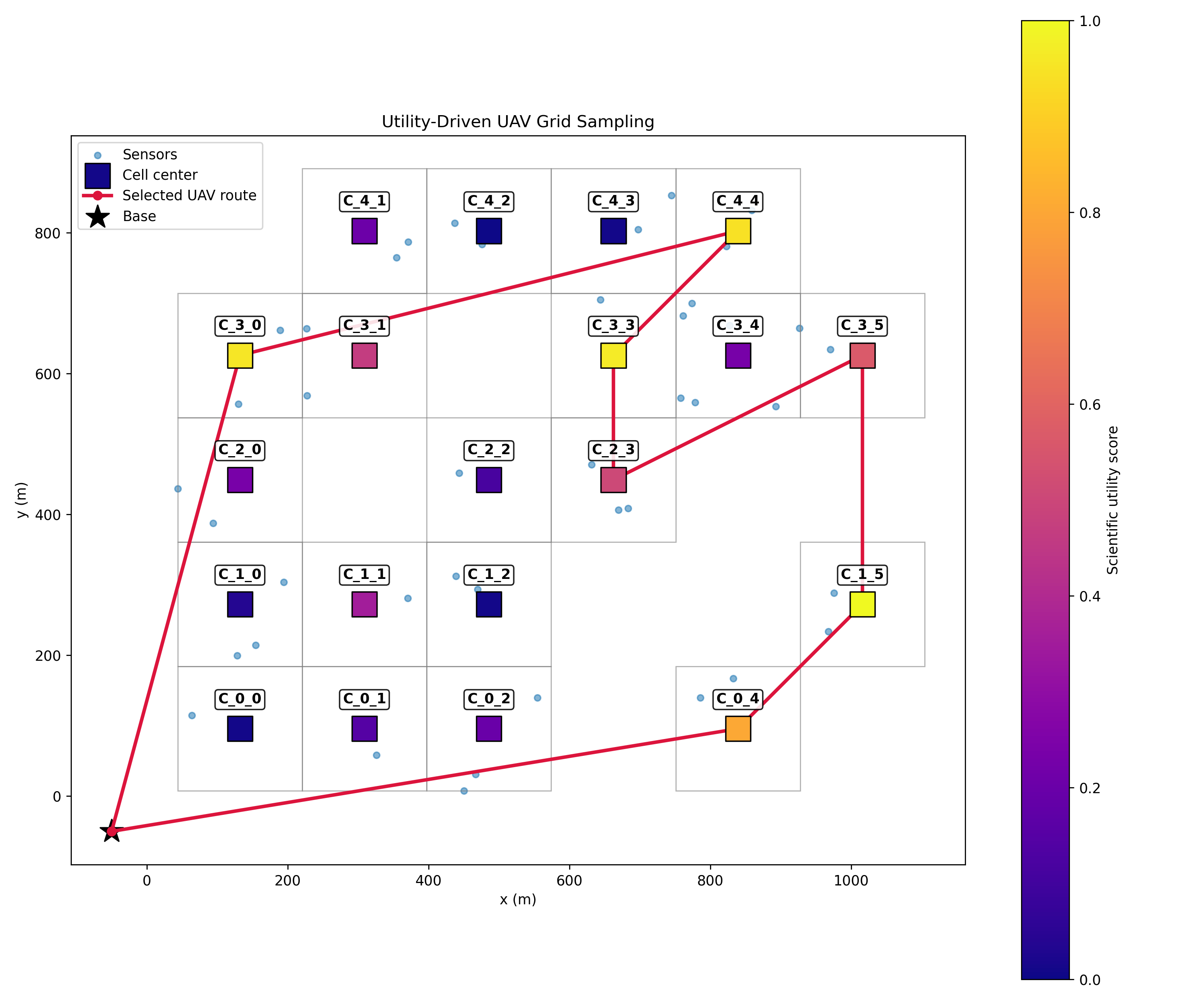}}
\vspace{-0.1in}
\caption{Utility-driven UAV sampling route over the semi-synthetic field deployment.} 
\label{fig:uav-route}
\vspace{-0.1in}
\end{figure}

We evaluated the framework using the field-scale soil-moisture sensor-network dataset under a semi-synthetic spatial deployment~\cite{dataset2017}. The field was divided into UAV-coverage cells, and each cell received a utility score based on freshness, anomaly likelihood, uncertainty, and redundancy.
Figure~\ref{fig:uav-route} shows the resulting utility-aware route. The UAV selected eight cells and returned safely to the base station. The total mission energy was 55,309.8 J, below the usable battery budget of 66,500 J, corresponding to 83.2\% budget utilization. The selected route prioritizes high-utility cells while accounting for travel, hovering, communication, and return-to-base energy. These preliminary results demonstrate the feasibility of utility-driven UAV sampling for energy-constrained smart-farming deployments.

These preliminary results demonstrate that the proposed framework can prioritize high-value cells while respecting the UAV energy constraint.

\bibliographystyle{ACM-Reference-Format}
\bibliography{ref}

@INPROCEEDINGS{soltani-smartfarm,
  author={Soltani, Keiwan and Tanwar, Vishesh Kumar and Gupta, Ashish and Das, Sajal K.},
  booktitle={2025 IEEE 22nd International Conference on Mobile Ad-Hoc and Smart Systems (MASS)}, 
  title={Energy-Efficient Split Learning for Resource-Constrained Environments: A Smart Farming Solution}, 
  year={2025},
  volume={},
  number={},
  pages={111-119},
  doi={10.1109/MASS66014.2025.00028}}

@article{UAV-agriculture2024,
  author  = {Komatineni, B. K. and Makam, S. and Meena, S. S.},
  title   = {A comprehensive review of the functionality and applications of unmanned aerial vehicles ({UAVs}) in the realm of agriculture},
  journal = {Journal of Electrical Systems and Information Technology},
  year    = {2024},
  volume  = {11},
  pages   = {57},
  doi     = {10.1186/s43067-024-00184-8}
}

@INPROCEEDINGS{soltani-data2024,
  author={Soltani, Keiwan and Coro, Federico and Das, Sajal K.},
  booktitle={2024 IEEE 21st International Conference on Mobile Ad-Hoc and Smart Systems (MASS)}, 
  title={Optimizing UAV-Assisted Data Collection in IoT Sensor Networks Using Dual Cluster Head Strategy}, 
  year={2024},
  volume={},
  number={},
  pages={279-287},
  doi={10.1109/MASS62177.2024.00045}}

@article{dataset2017,
author = "Caley Gasch and David Brown",
title = "{Data from: A field-scale sensor network data set for monitoring and modeling the spatial and temporal variation of soil moisture in a dryland agricultural field}",
year = "2017",
month = "4",
doi = "10.15482/USDA.ADC/1349683"
}
\end{document}